\documentclass{pnastwo}
\usepackage{epsfig,amsmath,amssymb,graphicx,color,calc,epstopdf}
\usepackage{dsfont}
\usepackage{color}

\let\a=\alpha  \let\g=\gamma \let\d=\delta
\let\e=\varepsilon \let\z=\zeta  \let\k=\kappa
\let\l=\lambda \let\m=\mu   
\let\s=\sigma   \let\c=\chi

\let\ee=\epsilon \let\r=\rho \let\th=\theta \let\io=\infty

  \def\OO{{\cal O}}

\def\to{\rightarrow} \def\la{\left\langle} \def\ra{\right\rangle}

\newcommand{\beq}{\begin{equation}}
\newcommand{\eeq}{\end{equation}}
\newcommand{\ba}{\begin{align}}
\newcommand{\ea}{\end{align}}

\contributor{Submitted to Proceedings
of the National Academy of Sciences of the United States of America}
\url{www.pnas.org/cgi/doi/10.1073/pnas.XXXXXXXXXX}
\copyrightyear{2014}
\issuedate{Issue Date}
\volume{Volume}
\issuenumber{Issue Number}

\begin{document}


\title{Universal Spectrum of Normal Modes in Low-Temperature Glasses: an Exact Solution }




\author{Silvio Franz\affil{1}{LPTMS, CNRS, Univ. Paris Sud, Universit\'e Paris-Saclay, 91405 Orsay, France},
Giorgio Parisi\affil{2}{Dipartimento di Fisica,
Sapienza Universit\`a di Roma, INFN, Sezione di Roma I, IPFC -- CNR, P.le A. Moro 2, I-00185 Roma, Italy},
Pierfrancesco Urbani\affil{3}{IPhT, CEA/DSM-CNRS/URA 2306, CEA Saclay, F-91191 Gif-sur-Yvette Cedex, France},
\and Francesco Zamponi\affil{4}{LPT,
\'Ecole Normale Sup\'erieure, UMR 8549 CNRS, 24 Rue Lhomond, 75005 France}}

\contributor{Submitted to Proceedings of the National Academy of Sciences
of the United States of America}

\maketitle

\begin{article}

\begin{abstract} 
We report an analytical study of the vibrational spectrum of the simplest model of jamming, 
the soft perceptron. We identify two distinct classes of soft modes. The first kind of modes are related to isostaticity
and appear only in the close vicinity of the jamming transition. The second kind of modes instead are present everywhere
in the glass phase and are related to the hierarchical structure of the potential energy landscape.
Our results highlight the universality of the spectrum of normal modes in disordered systems, and 
open the way towards a detailed analytical understanding of the vibrational spectrum of low-temperature glasses.
\end{abstract}






\noindent \noindent\rule{8cm}{0.4pt}
\\
{\bf Significance} 
The vibrational spectrum of glasses displays an anomalous excess of soft low-frequency modes
with respect to crystals. Such modes are responsible for many anomalies in thermodynamic
and transport properties of low-temperature glasses. Many distinct proposals have been formulated to understand
their origin but none of them results from the analytic solution of a microscopically grounded model.
Here we solve analytically the spectrum of a simple model that belongs to the same universality
class of glasses, and identify two distinct mechanisms that are responsible for the soft modes.

\noindent\rule{8cm}{0.4pt}

\paragraph*{Introduction --}
Low-energy excitations in disordered glassy systems
have received a lot of attention because of their multiple interesting features
and their importance for thermodynamic and transport properties of low-temperature glasses.
Much debate has concentrated around the deviation of the spectrum from the 
Debye law for solids, due to an
excess of low-energy excitations, known as the ``boson peak''~\cite{MS86}.  

The vibrational spectrum of glasses is a natural problem of
random matrix theory. In fact, the
Hessian of a disordered system is a random matrix due to the random position of particles in the
sample. The distribution of the particles induces non trivial
correlations between the matrix elements.  Many attempted to explain
the observed spectrum of eigenvalues by replacing the true statistical
ensemble with some simpler ones, in which correlations are neglected or
treated in approximate ways~\cite{SDG98,GMPV03,SRS07,XWLN07,Wy10,Za11,ML13,AKVOI13,Pa14,DLDLW14}.
Yet most of these
models are not microscopically grounded,
thus making it difficult to asses which of the proposed mechanisms are the most 
relevant and understand their interplay.

In this work we will focus on two ways of inducing a boson peak in random matrix models.
First, it has been
suggested that the boson peak is due to the
vicinity to the jamming transition where glasses are {\it isostatic}~\cite{OSLN03,LNSW10}.
Isostaticity means that the number of degrees of freedom
is exactly equal to the number of interactions.
Isostaticity implies {\it marginal mechanical
stability} (MMS): cutting one particle contact induces an unstable soft mode, that allows particles
to slide without paying any energy cost~\cite{WSNW05,Wy12}.
From this hypothesis, scaling
laws have been derived that characterise the spectrum as a function of
the distance from an isostatic point~\cite{DLDLW14,DLBW14,OSLN03}. 
Secondly, it has been proposed that low-temperature glasses have a complex energy
landscape with a hierarchical distribution of energy minima and barriers~\cite{nature}. Minima
are marginally stable~\cite{MPV87} and display anomalous soft modes~\cite{XVLN10,DLDLW14} related to the
lowest energy barriers~\cite{BW09b,ML11,Procaccia}. 
We will denote this second kind of marginality as {\it landscape marginal stability}~(LMS).

Both mechanisms described above are {\it highly universal}. LMS is a generic property of mean field strongly
disordered models~\cite{MPV87}. MMS holds for a broad class of simple random matrix models~\cite{Wy10,Pa14,DLDLW14,DLBW14} 
and for realistic glass 
models~\cite{OSLN03,MKK08,IBB12} at the isostatic point.
Universality motivates the introduction of a broad class of
Continuous Constraint Satisfaction
Problems (CCSP)~\cite{FP15}, in which a set of constraints is imposed on a set of continuous variables.
In the satisfiable (SAT) phase, all the constraints can be satisfied, while this is impossible in the unsatisfiable (UNSAT) phase.
A sharp SAT-UNSAT transition separates the two phases: jamming can be seen as a particular instance of this transition.
In fact,
{\it (i)} jamming properties are 
within numerical precision {\it super-universal}, i.e. independent of the spatial dimension $d$ for all $d\ge 2$~\cite{GLN12,CCPZ15},
{\it (ii)}~they can be analytically predicted through the exact solution in $d\to\io$~\cite{CKPUZ14,nature},
and {\it (iii)} the perceptron model of neural networks, a prototypical CCSP, displays a 
jamming transition with the same exponents~\cite{FP15}.
Based on universality, both for
analytical and numerical computations, the perceptron appears to
be the simplest model\footnote{There is of course the possibility of a weak
  dependence of the jamming exponents on $d$. In that
  case our results would be exact only for $d\to\io$, yet they can
  be expected to provide a very good approximation in $d<\infty$.} 
where low-temperature glassy behaviour
can be studied~\cite{FP15}.

Here, we exploit this simplicity and
characterise analytically the vibrational spectrum of the perceptron at zero temperature in the
glass phase. 
Our main results are: {\it (i)}
the spectrum is given by a Marchenko-Pastur law
with parameters that can be computed analytically;
{\it (ii)} it closely resembles the one
of soft sphere glass models in all $d \geq 2$;
{\it (iii)} it displays soft modes
coming from marginal stabilities of both kinds (LMS and MMS), allowing us to unify
both contributions and understand their interplay.
Our results are based on the replica method and random matrix theory, and for the first
time we are able to derive all the critical properties of
 jamming
 within the analytic solution
of a 
well-defined 
microscopic model.

\paragraph*{The model --}

We propose to use the perceptron as a minimal model for
  jamming. In doing that we heavily rely on a universality hypothesis.
  Rather then looking for physical realism, we posit that we can
  capture many of the interesting features of low-energy excitations
  in the glass phase close to jamming making
  abstraction of the following three basic properties of particle systems: 
  (a)~the relevant degrees of
  freedom, the particle positions, are continuous variables;
  (b)~in hard spheres, impenetrability can be seen as a set of constraints
  -inequalities- on the distances between spheres; (c)~spheres can be made soft by relaxing the
  impenetrability constraint and imposing a harmonic energy cost to
  any overlaps~\cite{OSLN03}.  

Let us now introduce a general class of CCSP where a
set of $N$ continuous variables $x=\{x_1,...,x_N\}$ is subject to a set
of $M$ constraints of the form $h_\mu(x) > 0$ ($\mu=1\cdots M$).
The ``hard'' version of the problem corresponds to allowing only configurations
that satisfy the contraints; the ``soft'' version corresponds to giving an energetic
penalty to each violated constraint. This can be encoded in an energy, or Hamiltionian, or ``cost'' 
function\footnote{Note that the choice of the exponent $2$ in Eq.~\eqref{eq:1}
is arbitrary but corresponds to the common choice of a soft harmonic repulsion
in the context of sphere packings; other exponents can be chosen and the results
remain qualitatively similar, see~\cite{OSLN03}.}
\beq
  \label{eq:1}
  H[x] =\frac \e 2 \sum_{\mu=1}^M h_\mu^2 \theta(-h_\mu) \ . \hskip20pt
  \eeq 
where $\th(x)$ is the Heaviside function. 
For all configurations $x$, one obviously has $H[x]\ge 0$.
There are thus only two possibilities: either
 all the constraints can be
  satisfied and the ground state energy is $H_{GS} =0$ (SAT phase), or
   $H_{GS} >0$ (UNSAT phase). These
   two phases are separated, in the thermodynamic limit $N\to\io$, by a sharp
  SAT-UNSAT phase transition~\cite{MM09}.
  The hard case corresponds to $\e=\io$, and the UNSAT phase is then forbidden.

Particle systems correspond to a special choice: the $x_i$ are $d$-dimensional vectors 
confined in a finite fixed volume; 
each constraint is the ``gap'' between two given particles, so it has the form $h_\mu = | x_i - x_j | - \s$,
where $\s$ is the particle diameter; the index $\mu = \{ i < j \}$ takes $M = N(N-1)/2$
values corresponding to all possible particle pairs. Plugging this in Eq.~\eqref{eq:1} the reader
will immediately recognize the soft sphere Hamiltonian used in most studies on jamming~\cite{OSLN03,JBZ11}.
Because jamming is the point where the energy first becomes non-zero upon increasing $\s$, 
we can identify it with the SAT-UNSAT transition for this special choice of the constraints.

The spherical perceptron is probably the simplest abstract CCSP 
where, appropriately rephrased, the three ingredients above are combined~\cite{FP15}.
The variables
$x$ belong to the unitary $N$-dimensional sphere with $\sum_i
x_i^2=1$, and one considers $M=\alpha N$ constraints of the form
\beq
  \label{eq:1a}
h_\mu=\xi^\mu\cdot x-\sigma>0 \ ,
\eeq
defined in terms of vectors,  $\xi^\mu=\{\xi_1^\mu,...,\xi^\mu_N\}$ composed by
quenched i.i.d. random variables with independent $N(0,1)$
components.
 The control parameters of the system are thus $\alpha$ and $\sigma$, and
jamming defines a line in the ($\alpha$,
  $\sigma$) plane separating the SAT and UNSAT phases. The sign of $\sigma$ is crucial:
  for $\sigma>0$ the perceptron is a 
convex CCSP, with a unique energy minimum~\cite{Ga88,GD88}; for $\sigma<0$ the problem
is non-convex and multiple minima are possible \cite{FP15}. 

In the following, $\la \bullet \ra$ indicates an average on the
minimum energy configurations for a given realisation of quenched
disorder (i.e. of the $\xi^\mu$) while $\overline{ \la \bullet \ra}$
indicates an additional average over the disorder.  We will also
introduce a special notation for ($\alpha$ times) the average moments
of the gap distribution in a given configuration, 
\beq\label{eq:rn}
[h^n] = \frac 1 N \sum_{\mu=1}^M h_\mu^n \theta(-h_\mu) \ .  
\eeq 

For future reference, it is useful to provide a simple dictionary
between physical quantities in particles systems and in the perceptron.
  The {\bf energy} is clearly identified with $H \propto [h^2]$ in both cases.
 The {\bf pressure} is proportional to
  $\partial H/\partial\sigma \propto [h]$ in particle systems, and the same definition can be used
  for the perceptron.  
The {\bf gaps} between pair of particles correspond to
  the constraints $h_ \mu$. 
  The {\bf forces}, which act from a
  constraint $\mu$ to a variable $i$, are naturally defined as the
  $\mu$-contribution to the total force $F_i = -\partial H/\partial x_i$
  acting on $x_i$, namely, $f_i^\mu=- (\partial h_\mu/\partial x_i ) h_\mu
  \theta(-h_\mu)$.
   The {\bf total number of
    contacts} in spheres is the number $N [1]$ of violated constraints with $h_\mu \leq 0$; one can keep the same definition
 for the perceptron.
  In the following we will approach jamming from the UNSAT phase, where
  $[h^n]\to 0$ from non-zero values\footnote{All the moments $[h^n]$,
      including $n=0$, are clearly equal to zero in the SAT phase.} for all $n>0$, while, for
  continuity, $[1]$ tends to the fraction of binding
  constraints, i.e. those such that $h_\mu=0$.
  The {\bf isostaticity} condition is that the number of binding
  constraints equals the number of degrees of freedom and can be
  therefore written as $[1]=1$. As already mentioned, $\alpha$ and $\sigma$ 
  play the role of control
  parameters that are analogue to the {\bf packing fraction} in the
  sphere problem.  In addition the {\bf Debye-Waller factor} 
corresponds to the  Edwards-Anderson parameter (see below and~\cite{CKPUZ14}). 
Finally, note that {\bf rattlers}, i.e. particles that at jamming are involved only in non-binding constraints,
cannot exist in the perceptron, because each variable $x_i$ is connected to all the constraints.

\paragraph*{Vibrational spectrum --}
We now present our main technical result which is the exact computation
of the eigenvalue spectrum of the Hessian of $H[x]$ in its points of minimum
(we now choose $\e=1$).
We enforce 
the spherical constraint through a Lagrange multiplier $\z$ and
consider the modified Hamiltonian
$H_\z[x]=H[x]-\frac N 2 \z(x^2-1)$.
The first order minimization conditions read 
\begin{eqnarray}
  \label{eq:2}
\frac{\partial H_\z}{\partial x_i} =  \sum_{\mu=1}^M\xi_i^\mu h_\mu \theta(-h_\mu) -N \z x_i=0 \ .
\end{eqnarray}
Multiplying by $x_i$ and summing over $i$ we can obtain a relation
between $\z$ and the distribution of the gaps $h_\mu$ in the minimum, namely
$  \z= \frac 1 N \sum_{\mu} (h_\mu^2+\s h_\mu) \theta(-h_\mu) 
=[h^2]+\s [h] $.
The Hessian matrix, normalized with $N$ reads:
\begin{eqnarray}
  \label{eq:5}
  M_{ij}=\frac1N \frac{\partial^2 H_\z}{\partial x_i \partial x_j} =
  \frac 1 N \sum_{\mu=1}^M\xi_i^\mu \xi_j^\mu \theta(-h_\mu) - \z
  \delta_{ij} \ .
\end{eqnarray}
 Notice that in the SAT phase, all the gaps $h_\mu$ are positive
and both $\z$ and the elements of $M$ are trivially equal to zero.
We concentrate therefore on the UNSAT
phase, where there is a non vanishing fraction $[1]$ of negative gaps
$h_\mu$.

In principle in the point of minima of $H$, $\xi_i^\mu \xi_j^\mu$ and
$\theta(-h_\mu) $ that appear in Eq.~\eqref{eq:5} could be effectively
correlated, however to the leading order in large $N$ these
correlations can be neglected, because each $h_\mu$ is the sum of a
large number of $\xi^\mu_i$. 
 The matrix $M$ is thus equivalent to a random matrix
from a modified Wishart ensemble \cite{ABdF11}, with an effective
number of random contributions equal to $N[1]$ and a constant term
$\z$ added on the diagonal: \beq \label{eq:6} M_{ij} \sim \frac1N
\sum_{\mu=1}^{N[1]}\xi_i^\mu \xi_j^\mu - \z \delta_{ij} = [1]
W_{ij}-\z \delta_{ij} \ , \eeq where $W_{ij} = (N[1])^{-1}
\sum_{\mu=1}^{N[1]}\xi_i^\mu \xi_j^\mu$ is a standard Wishart matrix \cite{W28}
with ``quality factor'' $Q=1/[1]$.  It follows that for large $N$ the
eigenvalue spectrum of $M$ obeys the modified Marchenko-Pastur (MP)
law \cite{MP67} \beq
\begin{split}
  \label{eq:7}
 & \rho(\lambda)=
\left\{
\begin{array}{ll}
 (1-[1])\delta(\lambda+\z)+\nu(\lambda) & [1]<1 \\
 \nu(\lambda) & [1]>1
\end{array}
\right. \ ,
\\
&\nu(\lambda)=\frac{1}{2 \pi}
\frac{ \sqrt{ (\lambda-\lambda_{-})(\lambda_{+}-\lambda)}
}{\lambda+\z} \mathds{1}_{\lambda_-,\lambda_+}(\lambda) \ , \\
&\lambda_{\pm}= (\sqrt{[1]} \pm 1 )^2 - \z \ .  
\end{split}
\eeq
This result is very general: for any minimum of $H$, Eq. \eqref{eq:7} holds for specific
values of the parameters $[1]$ and $\z$.
Also, the eigenvectors of Wishart matrices are
delocalized~\cite{bai2007} and are asymptotically distributed according to the
uniform Haar measure on the sphere.
The same properties hold for the 
eigenvectors of the Hessian 
of the perceptron. 
We will see that Eq.~\eqref{eq:7} reproduces all the known features
of low-energy excitations close to jamming: its
main virtue is to relate these features to a
few characteristics of the gap distribution.

The condition of minimum of $H$ requires that all the eigenvalues of
the spectrum are positive or zero.  
For $[1]<1$, this implies $\zeta\leq 0$, which can
only happen if $\sigma>0$. Thus for $\sigma<0$, necessarily $[1]\geq 1$. In that case
we need $\lambda_{-}\ge 0$, i.e.
\begin{eqnarray}
  \label{eq:17}
  ( \sqrt{[1]} -1  )^2 \geq \z=[h^2]+\s [h]  \ . 
\end{eqnarray}
Equality in Eq.~\eqref{eq:17} corresponds to a {\it marginally
  stable} minimum whose spectrum touches zero. It also implies that at jamming, where $[h]$ and $[h^2]$
vanish and $[1]$ tends to the number of binding constraints,
marginally stable minima are necessarily isostatic with $[1]= 1$. 
Note that in the context of sphere packings, Eq.~\eqref{eq:17} translates
into a relation between the excess of contacts $\d z \sim [1]-1$ and
the pressure
$p \sim [h]$, which reads $\d z^2 \geq Const. \; \times\;  p$ and has been derived in~\cite{WSNW05}.

We now need to compute the moments $[h^n]$ that enter
in Eq.~\eqref{eq:7}.  Unfortunately, this computation cannot be done
analytically for a single minimum or a single sample. Instead, we are
able to compute the average $\overline{\la [h^n] \ra}$
over all the absolute minima of the Hamiltonian.
Because the moments 
$[h^n]$ are self-averaging for large $N$,
this provides information over the typical absolute minima of the Hamiltonian 
as a function of the control parameters $(\a,\s)$. 
In the following paragraphs we report the computation of $\overline{\la [h^n] \ra}$; for simplicity, unless otherwise
specified we drop the 
averages and replace $\overline{\la [h^n] \ra} \to [h^n]$.

 \begin{figure}[t]
\begin{center}
\includegraphics[width=.4 \textwidth]{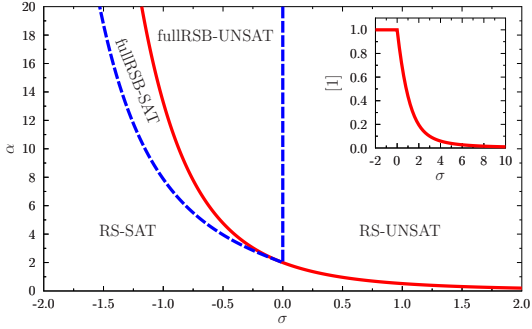}
\end{center}
\caption{(Main panel)
Phase diagram of the model. The RSB region is delimited by the dashed blue line, $\s<0$ and $\a>\a_c(\s)$. 
The jamming line $\a_J(\s)$ that separates the SAT from the UNSAT phase,
  as estimated by the RS solution Eq.~\eqref{eq:37}, is depicted as a full red line. 
The exact $\a_J(\s)$ should be computed within the fullRSB ansatz for $\s<0$,
but we expect only a small difference as confirmed in Fig.~\ref{fig:4}. 
(Inset) The {\it density of contacts} $[1]$ along the jamming
  line. Jamming is isostatic ($[1]=1$) for $\sigma \leq 0$ and
  hypostatic ($[1]<1$) for
  $\sigma>0$.
\label{fig:pd}
}
\end{figure}

\paragraph{Thermodynamic analysis: the convex domain --}
Thermodynamic and disorder averages can be computed with
the aid of the replica method~\cite{MPV87}.
The partition function is written as an integral over a certain number of copies of the system,
the average of the disorder is taken and the resulting integral is evaluated through the saddle-point
method for $N\to\io$. 
As a result, one should minimize
a free-energy which is a function of the average overlap between replicas,
$q_{ab} = \overline{\langle x^a \cdot x^b \rangle}$  ($q_{aa}=1$ due to the spherical constraint).
The minimization is not possible for a generic matrix $q_{ab}$ and one
has thus to make an {\it ansatz} on the structure of the matrix minimizer,
that codes for the organization
of the ergodic complonent in the system~\cite{MPV87}. If
ergodicity holds and there is a single component, the replica
symmetric (RS)  $q_{ab} = q$ for $a \neq b$
ansatz is appropriate.  
The RS free energy for the perceptron is~\cite{Ga88}
\begin{eqnarray}
  \label{eq:33}
&&  F_{RS}=-\frac T 2 \left[\log(1-q)+\frac{q}{1-q}\right]-\\
&&\alpha T\int
  D_q(h+\sigma)\log \left(\int
D_{1-q}(y-h) \ e^{-\beta y^2 \theta(-y)/2} \nonumber 
\right) \ , \end{eqnarray}
where $D_q(h) = dh \exp[-h^2/(2q)]/\sqrt{2\pi q}$ and $D(h) = D_1(h)$. This expression must be minimized with
respect to $q$.

We have two distinct situations when $T\to 0$.
{\it (i)} in the SAT phase, because there are many solutions, different replicas
can be in different solutions and $q<1$ at $T=0$~\cite{Ga88}: from Eq.~\eqref{eq:33} one
can show that the ground state energy is $E_0=0$.
{\it (ii)} in the UNSAT phase, the replicas 
are in the unique absolute energy minimum and $q=1$ at $T=0$. For $T\to 0$,
due to harmonic vibrations,
$q = 1 - \c T + \OO(T^2)$. 
The limit $T\to 0$ must therefore be taken with $q\to 1$ and
$\chi=\beta(1-q)$ fixed.
The parameter $\c$ diverges approaching the SAT-UNSAT transition from the UNSAT phase,
while $q\to1$ approaching the transition from the SAT phase.

Let us now focus on the UNSAT phase.
Evaluating the most internal integral by
saddle point one finds
\begin{eqnarray}
  \label{eq:34}
  E_0=\overline{\la H \ra}=-\frac{1}{2 \chi} +\frac {\alpha}{2 (1+\chi)}\int_{-\infty}^0 D(h+\sigma)h^2 \ ,
\end{eqnarray}
and optimizing over $\c$ gives the equation 
\begin{eqnarray}
  \label{eq:35}
  \left(1+\frac 1 \chi
\right)^2=\alpha \int_{-\infty}^0 D(h+\sigma)h^2 \ .
\end{eqnarray}
Also, one can show that 
\begin{eqnarray}
  \label{eq:36}
  [h^n]=\frac {\alpha}{ (1+\chi)^n}\int_{-\infty}^0 D(h+\sigma)h^n \ .
\end{eqnarray}
Jamming is the point for
which $\chi\to\infty$, or 
\begin{eqnarray}
  \label{eq:37}
  \alpha^{\rm RS}_J(\sigma)= \left( \int_{-\infty}^0 D(h+\sigma)h^2\right)^{-1} \ ,
\end{eqnarray}
which coincides with the result of~\cite{Ga88}.
Also, $1/\chi \propto \ee$ vanishes
linearly in the distance $\ee$ from the line $\alpha^{\rm RS}_J(\s)$, and
from Eqs.~\eqref{eq:35} and \eqref{eq:36} we obtain that
$E_0 = [h^2]/2 = 1/(2 \chi^2) \propto \ee^2$ vanishes quadratically~\cite{OSLN03,JBZ11}.
We can thus identify the line $\alpha^{\rm RS}_J(\s)$ with the jamming transition,
because $E_0 > 0$ for $\a > \alpha^{\rm RS}_J(\s)$ while $E_0 =0$ for
$\a \leq \alpha^{\rm RS}_J(\s)$.
From Eqs.~\eqref{eq:36} and \eqref{eq:37} we get $[1] <
1$ on the jamming line (Fig.~\ref{fig:pd}), where the system is thus hypostatic
and not critical~\cite{FP15}.

From Eq.~\eqref{eq:36} we can compute the moments that enter in 
Eq.~\eqref{eq:7}.  Recall that the RS solution implies a
unique minimum of the energy, so the
average $\overline{\la [h^n] \ra}$ coincide with the value of
$[h^n]$ in the absolute minimum.  In the UNSAT phase for $\s>0$ we get:
{\it (i)} for large $\a$, $[1]>1$ and $\l_->0$: the spectrum is gapped;
{\it (ii)} for $\a \gtrsim \a_J^{\rm RS}(\s)$, $\l_- > 0$ and $[1]<1$ with $\z<0$:
the spectrum is gapped and it has $1-[1]$ modes with eigenvalue $\lambda=-\zeta>0$;
{\it (iii)} the same remains true at jamming for $\s>0$ and $\a= \a_J^{\rm RS}(\s)$, except
that the $1-[1]$ modes vanish trivially because $\z=0$.
When $\s \to 0$,
$\lambda_- \propto \sigma^2$ vanishes so the gap closes on the line $\s=0$.

Finally, one can
study the stability condition of the RS solution by considering a small
perturbation of the matrix $q_{ab}$ and checking if this perturbation
lowers the free energy. A standard computation~\cite{AT78,GD88} leads to the stability condition.
In the SAT phase where $q<1$, the RS stability condition is $\a \leq \a_c(\s)$ as computed in~\cite{FP15};
the line $\a_c(\s)$ falls in the SAT region for $\s\leq 0$.
In the UNSAT phase, we get the condition
\beq
  \label{eq:9}
   \left(1+\frac 1 \chi \right)^2 \geq [1] \hskip10pt \Leftrightarrow \hskip10pt \int_{-\infty}^0 D(h+\sigma)(h^2-1) \geq 0 \ ,
\eeq
which is verified for $\sigma \geq 0$ while it is violated for $\s<0$. 

In summary,
for $\s < 0$ and $\a > \a_c(\s)$ the RS solution is unstable (as depicted in Fig.~\ref{fig:pd}) and
one must consider a solution for $q_{ab}$ that is not invariant under permutation symmetry,
usually denoted a {\it replica symmetry breaking} RSB solution~\cite{MPV87}.
Where the RS solution is correct, the spectrum is gapped.

\paragraph{Thermodynamic analysis: the non-convex domain --}
In the region of non convex optimization $\s<0$, ergodicity is broken at low
temperatures and large $\a$. The RS solution is unstable and the structure
of the matrix $q_{ab}$ is parametrized by a function $q(x)$ defined in
the interval $x\in[0,1]$, which encodes the values
of the overlaps of replicas that populate different metastable states
of the system. This is called a {\it full replica symmetry
  breaking} (fullRSB) ansatz~\cite{MPV87}. 
   In particular this implies that at $T=0$ there are many quasi-degenerate
minima of the Hamiltonian.  The value of $q(1) = q_{\rm
  EA}$ is the Edwards-Anderson order parameter
that describes the overlap of replicas confined in the same metastable
state.  The fullRSB equations for the perceptron have been written
e.g. in~\cite{GR97}. In general they can only be solved numerically,
but the scaling around the jamming transition can be obtained
analytically~\cite{CKPUZ14,nature,FP15}. Here we discuss the main
results of this analysis, a detailed derivation will be reported
elsewhere. 

As in the RS case, $q_{\rm EA} = 1 - \c T$ in the UNSAT phase with $\c \to \io$ at the jamming transition.
 The jamming line falls in the fullRSB region for all $\s<0$ (Fig.~\ref{fig:pd}) and can thus be
computed numerically solving the fullRSB equations at $T=0$; however, we expect a small difference with the RS computation, and
in general the RS result is an upper bound, $\a^{\rm fullRSB}_J(\s) < \a^{\rm RS}_J(\s)$, so we did not perform the fullRSB 
computation.
Let us call once again $\ee$ the distance from the jamming line.
Combining the results of~\cite{CKPUZ14,nature,FP15} with original results derived in this work, 
the following properties can be obtained analytically for $\ee \to 0$:
\begin{itemize}
\item[{\it (a)}] the system is isostatic with $[1]=1$ identically for all $\s<0$ and $\a = \a^{\rm fullRSB}_J(\s)$;
\item[{\it (b)}] in the UNSAT phase $\c \sim \ee^{-1/2}$; the average
energy vanishes at jamming as $\overline{\la H \ra} \propto [h^2] \propto \ee^2$; the average gap is $[h]\propto \ee$; and the excess
of contacts above the isostatic value is $[1] -1  \sim \ee^{1/2}$;
\item[{\it (c)}] in the SAT phase, the Edwards-Anderson order parameter behaves
as $1 -q_{\rm EA} \sim \ee^{\kappa}$;
\item[{\it (d)}] at jamming ($\ee=0$) the probability distribution of the gaps, defined as $g(h) = \text{Prob}(h^\m=h)$ 
satisfies $g(h) \sim h^{-\g}$ for $h\to 0^+$, while the distribution
of absolute values of the forces 
satisfies $P(f) \sim f^\theta$ for $f\to 0^+$; 
\item[{\it (e)}] the values of the critical exponents $\k=1.41574\ldots, \ \g=0.41269\ldots, \ \th=0.42311\ldots$ 
are obtained analytically and coincide with the ones of soft spheres
in mean field.
\end{itemize}
We can next focus on the spectrum and compute $\l_-$ and $\z$ that appear in Eq.~\eqref{eq:7}. We obtain that
\begin{itemize}
\item[{\it (f)}] $\l_-$ vanishes identically in the fullRSB phase, because the condition $\l_-=0$ coincides with the LMS
condition\footnote{In technical terms, $\l_-=0$ is equivalent to the vanishing of the {\it replicon} mode of the fullRSB free energy.}.
Hence, in the fullRSB UNSAT phase, the spectrum is $\r(\l) \sim \sqrt{\l}/(\l + \z)$ for small $\l$ and energy minima are marginally stable.
\item[{\it (g)}] $\z$ is positive in the UNSAT phase but it goes to zero, as expected, at the jamming transition. Therefore, at jamming
$\r(\l) \sim 1/\sqrt{\l}$ has a much larger density of soft modes. Slightly away from jamming, $\r(\l) \sim \sqrt{\l}$ for $\l \ll \z$, then reaches
a maximum $\r(\l \sim \z) \sim 1/\sqrt{\z}$, and then decreases as $\r(\l) \sim 1/\sqrt{\l}$ for $\l \gg \z$.
\end{itemize}
We thus identify two distinct contributions to soft modes: the fullRSB structure (LMS) induces marginality with $\r(\l) \sim \sqrt{\l}$ while
the proximity to jamming (MMS) induces a much stronger contribution with $\r(\l) \sim 1/\sqrt{\l}$.

Note that
although each of points {\it (a)-(g)} has been derived separately through scaling arguments or 
numerically~\cite{OSLN03,WSNW05,BW09b,LNSW10,Wy12,DLBW14,DLDLW14,nature,CKPUZ14}, here for the first time
we derive all of them in a unified way from the analytical solution of a well defined microscopic model.

\paragraph{Comparison with numerical data --} 
The results of the previous section can be compared with the numerical
minimization of the soft perceptron Hamiltonian given in Eq.~\eqref{eq:1}. 
To obtain numerically the minima of $H$, we use the following procedure.
We start from a random assignment $x_0$ of the variables and we use the routine
{\tt gsl\_multimin\_fdfminimizer\_vector\_bfgs2} of the GSL library~\cite{GSL}, that uses the
vector Broyden-Fletcher-Goldfarb-Shanno (BFGS) algorithm
to minimize a function. To implement the spherical constraint, we minimize
$H_A[x] = H[x/|x|] + A(x^2 - 1)^2$, where $A$ is an irrelevant number of order 1.
Note that this algorithm produces local minima that do not necessarily coincide with the
absolute ones in the non-convex domain $\sigma<0$. Therefore, the equilibrium
calculation should not necessarily provide exact results for the minima that we produce
numerically,  however, the differences -if any-
  between the minima found numerically and the theoretical
  expectations for the absolute ones are very
  small. This is probably due to the fact that we work in a regime of
  fullRSB where relevant metastable states have energy {\it
    density} equal to the one of the ground state \cite{MPV87}.
    
    \begin{figure}[t]
\begin{center}
\includegraphics[width=.4 \textwidth]{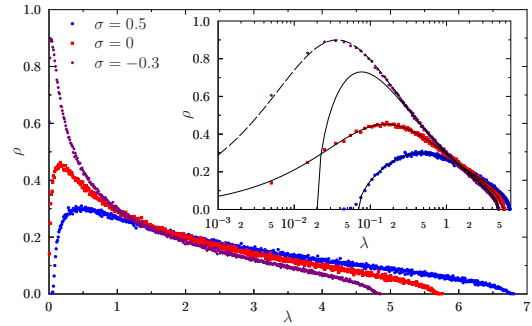}
\end{center}
\caption{Spectrum of the Hessian for $N=1600$, $\alpha=4$ and
  $\sigma=0.5,0,-0.3$, averaged over $208$ samples, in linear (main panel) and semilog (inset) scales.
  The MP law given in Eq.~\eqref{eq:7}
  with RS parameters (full lines)
  perfectly reproduces the data for $\s\geq 0$, while deviations are observed for
  $\sigma<0$, where instead a MP law with $\z=0.037$ and
$[1]=(1+\sqrt{\z})^2 =1.42$ (hence $\lambda_-=0$ and $\lambda_+=4.76$) 
perfectly fits the spectrum (dashed line).
  \label{fig:2} }
\end{figure}

\begin{figure}[t]
\begin{center}
\includegraphics[width=.4 \textwidth]{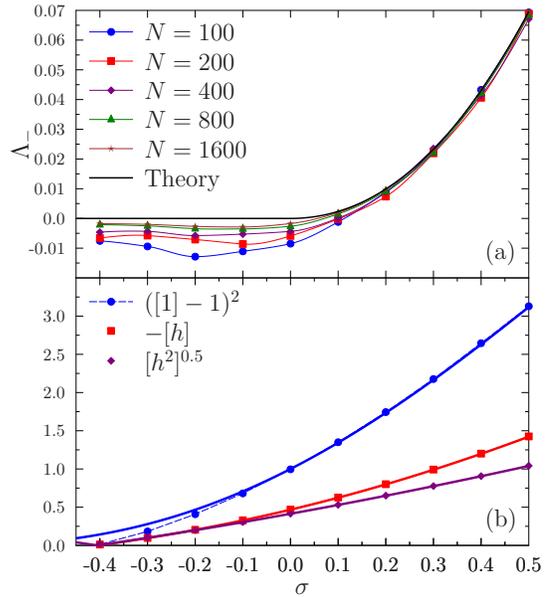}
\end{center}
\caption{
(a)
The combination   $\Lambda_- = (\sqrt{[1]}-1)^2-([h^2]+\sigma [h])$
  as a function of $\sigma$ for $\alpha=4$ and several $N$, averaged over $100$ samples. 
  In the thermodynamic limit this tends to the edge of the
  spectrum of the Hessian. 
 Stability
  requires $\Lambda_-\ge 0$ for $N\to\infty$. For $\sigma>0$,
  $\Lambda_-$ tends to values greater than 0 that coincide with the analytical RS result (full line). 
  For $\sigma<0$ the
  data indicate marginal stability, $\Lambda_-\to 0$ for $N\to\infty$.
(b)  The moments $([1]-1)^2$, $[h]$ and $\sqrt{[h^2]}$ as
  a function of $\sigma$ for $\a=4,N=1600$ (points with dashed lines as guides to the eye).
  Full lines are the RS predictions.
While deviations between theory and simulations 
are observed in the behaviour of $[1]$, there are
no appreciable differences for $[h]$ and $[h^2]$.
As predicted by the theory these quantities vanish linearly at the
    jamming point estimated here at $\sigma_J\approx -0.409$, which is very
    well approximated by $\sigma_J^{\rm RS} = -0.405234$. 
    \label{fig:3} \label{fig:4} 
 }
\end{figure}

In Fig.~\ref{fig:2} we report the spectrum computed numerically and we compare it
with the theoretical prediction. As expected, in the RS phase the absolute
minimum is unique and can be easily found numerically. Hence, the theoretical
prediction perfectly coincides with the analytical result. On the contrary, in the fullRSB
phase the numerical algorithm gets stuck into local minima.
Yet even in this case the spectrum is described by a Marchenko-Pastur law, which confirms
that the result in Eq.~\eqref{eq:7} holds for all minima. Moreover, we find $\l_-=0$ suggesting that
the local minima found by the algorithm are marginally stable.

We checked the expected delocalization properties of the
eigenvectors of the Hessian through the statistics of the
inverse participation ratio, 
that for a given normalized eigenvector $v_i$ 
is defined as 
$  y=\sum_{i=1}^N v_i^4$, 
and is of order $O(1/N)$ for delocalized eigenvectors. We found that the distribution of $y$ is independent of $\alpha$
and $\sigma$. Moreover, the average value of $Ny$ tends to the value $3$
implied by the Haar distribution on the sphere.

In Fig.~\ref{fig:3} we report the value of $\Lambda_- =
(\sqrt{[1]}-1)^2-([h^2]+\sigma [h])$, where the moments $[h^n]$ are
evaluated on the numerically obtained minima. According to
Eq.~\eqref{eq:7}, this quantity should tend to the value of the edge
of the spectrum 
$\lambda_-$ in the
thermodynamic limit $N\to\io$. As expected, we observe that $\Lambda_-$ is
positive and follows the analytic RS prediction for $\s>0$, while for
$\s<0$ we observe that $\Lambda_-\to 0$ for $N\to\io$.
Also, in Fig.~\ref{fig:4} we report the moments $([1]-1)^2$, $[h]$ and $\sqrt{[h^2]}$ as a function of $\s$
in the UNSAT phase. As predicted by the scaling analysis of the fullRSB solution, these quantities vanish linearly
in $\ee = \s - \s_J$ where $\s_J$ is the jamming point. We see
therefore that all the regimes predicted by the theory are observed in
numerical simulations.

\paragraph*{Characteristic frequencies and the boson peak --}
We now show that defining the frequency $\omega=\sqrt{\lambda}$
and the density of states $D(\omega)=\rho(\lambda)\frac{d\lambda}{d\omega}$,
our spectrum reproduces the salient features of the boson peak phenomenology as described in the introduction.
Following~\cite{DLDLW14},
we define the characteristic frequencies $\omega_* = \sqrt{\z}$, $\omega_0 = \sqrt{\l_-}$ and $\omega_{\rm max} = \sqrt{\l_+}$,
and from Eq.~\eqref{eq:7} we obtain
\begin{eqnarray}
  \label{eq:8}
D(\omega)
=\frac{1}{ \pi}
\frac{ \omega \sqrt{ (\omega^2-\omega_0^2)(\omega_{\rm max}^2-\omega^2)}
}{\omega^2+\omega_*^2} \ ,
\hskip10pt \omega_0 \leq \omega \leq \omega_{\rm max} \ .
\end{eqnarray}
In the RS phase, we have $\omega_0 > 0$ and the spectrum is gapped, but
in a $d$-dimensional, translationally
invariant system one should see a Debye spectrum $D(\omega) \sim \omega^{d-1}$ for
$\omega<\omega_0$. 
For small $\omega_0 \ll \omega_*$, following~\cite{DLDLW14}, one expects
in dimension $d$:
\beq
D(\omega) \sim \begin{cases}
\omega^{d-1} & \omega \ll \omega_0 \ , \\
\omega^2/\omega_*^2     & \omega_0 \ll \omega \ll \omega_* \ , \\
\text{flat}     & \omega_* \ll \omega \ll \omega_{\rm max} \ ,
\end{cases}
\eeq
the phononic regime being absent for $d\to\io$ and in the perceptron. For $d=3$,
$D(\omega)$ displays a crossover between two $\omega^2$ regimes at $\omega_0$, the second having
a larger prefactor because $\omega_*$ is small, which results in a boson peak~\cite{DLDLW14}.

In the LMS (fullRSB) phase, $\lambda_{-}=0$ and thus $\omega_0=0$ identically.
We get
\beq\label{eq:17111}
D(\omega) =\frac{1}{ \pi}
\frac{ \omega^2 \sqrt{ \omega_{\rm max}^2-\omega^2}
}{\omega^2+\omega_*^2} 
\sim \begin{cases}
 \omega^2  &\text{for } \omega \ll \omega_* \ , \\
 \text{flat} &\text{for } \omega_* \lesssim \omega \leq \omega_{\rm max}   \\
\end{cases} 
\eeq
and the phonons should be completely hidden by the soft LMS excitations.
Here $\omega_* >0$ away from jamming, while
 $\omega_* = \sqrt{\z} \sim \sqrt{\s [h]} \sim \sqrt{\ee}$ when the distance from jamming $\ee$ goes to zero.

The result in~\eqref{eq:17111} is fully consistent with the boson peak anomaly in the excitation spectrum of soft sphere packings 
as known from simulations and scaling arguments~\cite{OSLN03,WSNW05,SLN05,MKK08,IBB12,DLDLW14},
confirming the super-universal behavior of glassy systems close to jamming. 
They are also fully consistent with the results of~\cite{DLDLW14}, with the advantage that here we can obtain a fully microscopic
expression of the characteristic frequencies $\omega_0$ and $\omega_*$. 

\paragraph*{Conclusions --}
The soft perceptron is simple enough to allow for a fully analytic
characterization of vibrational spectra around UNSAT energy minima.
While for any minimum of the Hamiltonian the spectrum has the form of
a Marchenko-Pastur law, here we computed the parameters of this distribution
only for the absolute minima.  Super-universality of the jamming
transition allows us to hypothesize that the predicted form and
parameter evolution of the spectrum captures many of the low energy
features of the spectrum of soft-sphere systems.  We find two kinds of
soft excitations, as described in~\cite{DLDLW14}.  The first ones are
related to the existence of a complex energy landscape characterized
by a multiplicity of quasi-degenerate marginally stable minima.  Due
to this landscape marginal stability (LMS), the low-energy spectrum is
$D(\omega) \sim \omega^2$.  The second ones are related to the
proximity to an isostatic jamming point, where the spectrum is
$D(\omega) \sim \omega^2/(\omega^2 + \ee)$. Hence, above a typical
frequency $\omega_*\sim \ee^{1/2}$ the density of states is flat, as
found in soft spheres~\cite{OSLN03,WSNW05}.
Note that in particle systems the Debye contribution of phonons to the spectrum scales as $\omega^{d-1}$, 
hence as soon as $d>3$ the contribution of the LMS soft modes should overcome the one of phonons. In $d=3$
the two contributions are of the same order, and therefore LMS modes might be mixed with phonons. Yet we think that LMS might explain
why anomalous soft modes distinct from phonons, that are responsible for the plasticity of the glass, 
are observed in three dimensional soft sphere packings in the jammed phase
for low frequencies $\omega < \omega_*$ in the boson peak region~\cite{WSNW05,BW09b,XVLN10,ML11,DLDLW14,Procaccia}.
A way to test this idea would be to compute numerically the vibrational spectrum of soft spheres
in $d\geq 3$ and investigate the evolution with $d$. Localized soft modes 
(e.g. the buckling modes discussed in~\cite{DLBW14,CCPZ15})
are also likely to emerge in finite dimensional systems and complicate the analysis.

Let us stress once again that all of our results have been obtained 
analytically through the exact solution of a well defined model, and there is hope that they will be derived in a mathematically rigorous way 
in the future.
Beyond their direct relevance for the physics of jamming, our results also open a new connection between jamming/packing problems
and constraint satisfaction problems with continuous variables, that we conjecture to display a SAT-UNSAT transition in the same
(super-)universality class of jamming~\cite{FP15}.

The analysis can be extended in several directions. First of all, one can study the spectrum in the SAT (unjammed) phase, corresponding
to hard spheres, despite the fact that 
the energy is zero. In fact, one can study the problem at finite temperature using the Thouless-Anderson-Palmer approach~\cite{TAP77,Me89} 
and then take the limit $T\to 0$ by properly scaling the frequencies~\cite{BW09b,MKK08,IBB12,DLBW14}. This computation would provide
an elegant analytic approach to reproduce the experimental results obtained for colloids in~\cite{GCSKB10,CEZetal10}.
Other quite straightforward extensions could be the study of the statistics of avalanches~\cite{LMW10}, and the study of a ``quantum
perceptron" to investigate how LMS affects the thermodynamic properties in the quantum regime, which could shed light on the mechanisms
that induce tunnelling two-level systems in glasses~\cite{AHV72}.


\paragraph*{Acknowledgments --}
The European Research Council has provided financial support 
through ERC grant agreement
no.~247328 and NPRGGLASS. 
We thank G.~Ben Arous, P.~Charbonneau, E.~Corwin, A.~Liu, L.~Manning, S.~Majumdar, 
A.~Poncet, C.~Rainone, G.~Schehr, P.~Vivo for very useful discussions.
We especially thank E.~Degiuli and M.~Wyart for pointing out the relation with~\cite{DLDLW14} and 
detecting an error in a preliminary version of the manuscript, and C.~Goodrich for providing
numerical data for soft disk packings~\cite{GoPhD}.

\vskip-10pt


\end{article}
\end{document}